\documentclass[12pt, prd]{revtex4}
\usepackage{amssymb}
\usepackage{amsmath}
\usepackage{graphicx}

\begin{document}

\title{Kinematic censorship and high energy particle collisions in the
Schwarzschild background}
\author{A. V. Toporensky}
\email{atopor@rambler.ru}
\affiliation{Sternberg Astronomical Institute, Lomonosov Moscow State University, 119991
Moscow, Russia}
\author{O. B. Zaslavskii}
\email{zaslav@ukr.net}
\affiliation{Department of Physics and Technology, Kharkov V.N. Karazin National
University, 4 Svoboda Square, Kharkov 61022, Ukraine}

\begin{abstract}
We consider near-horizon collisions between two particles moving freely in
the Schwarzschild metric in the region outside the horizon. One of them
emerges from a white \ hole. We scrutiny when such a process can lead to the
indefinitely large growth of the energy in the center of mass frame in the
point of collision. We also trace how the kinematics of collision manifests
itself in preserving the principle of kinematic censorship. According to
this principle, the energy released in any event of collision, must remain
finite although it can be made as large as one likes. Also, we find that
particle decay near the singularity leads to unbounded release of energy
independently of its initial value.
\end{abstract}

\maketitle

\section{Introduction}

High-energy particle collisions near black holes became a hot topic after an
observation made in \cite{ban}. If two particles fall towards the rotating
black hole and collide near its horizon, under certain conditions the energy 
$E_{c.m.}$ in the center of mass frame can grow unbounded. This is called
the BSW effect (after the names of its authors). It was found for extremal
rotating black holes and requires one of particles to be fine-tuned, $%
E=\omega _{H}L$ \cite{ban}, \cite{prd}. Here, $E$ is the Killing energy, $L$
being the angular momentum, $\omega _{H}$ the angular speed of a black hole.
For the Schwarzschild black hole this is impossible since $\omega _{H}=0$.
Meanwhile, there is another version (or analogue) of the BSW effect, valid
even in the Schwarzschild background. It implies head-on collision between
particles, when, say, particle 1 approaches the horizon while particle 2
moves away from \ it. In doing so, no fine-tuning is required, $E_{c.m.}$
grows indefinitely for any individual energies. As a particle cannot emerge
from the black hole horizon, this scenario implies the presence of a white
hole. Collisions of such a type were considered in \cite{gpwhite}, \cite%
{white}. Before, the fact that collision between particles with opposite
radial velocity leads to unbounded $E_{c.m.}$ was noticed in \cite{pir1} - 
\cite{pir3}. However, it was not pointed out there that any reasonable
physical realization of such a scenario corresponds just to white (not
black) hole horizons. A coherent scheme of possible high energy particle
collisions involving particles that emerge from a white hole, was built in 
\cite{frontal} depending on possible type of trajectory.

Meanwhile, in the present article, we consider other aspects of head-on
collisions. We take a simplest and physically relevant metric - the
Schwarzschild one. Then, all particles are not fine-tuned, so we consider
collisions between two usual particles. This corresponds to line 1 in Table
1 of Ref. \cite{frontal}. Thus we restrict ourselves by one type of
collision only but for this type scrutinize all possible types of
trajectories in space-time and their kinematic properties. In this context,
we make special emphasis on manifestation of the principle of kinematic
censorship \cite{cens}. According to this principle, in any event of
collision, only finite energy can be released. If some scenario admits
literally infinite $E_{c.m.}$, it is based on some unphysical assumption or
mistake and should be rejected. Sometimes, it is quite non-trivial task to
trace how this principle reveals itself in concrete scenarios \cite{gen}. In
the present work, we take a simplest case and consider collisions in "our"
region (outside the horizon). Even in this case, corresponding explanation
how this principle reveals itself, is quite involved and relies heavily on
some subtle details of particle kinematics.

One reservation is necessary. Although the fine-tuned trajectories in common
sense do not exist in the Schwarzschild space-time, there exist very special
scenarios in \ which a particle with small, although nonzero energy $%
E\approx 0$ plays a crucial role \cite{gp11} - \cite{immov}. We will not
discuss this degenerate scenario in the present paper.

The notion of white \ holes was introduced in \cite{delay} where it was
conjectured that they can be realized as some kinds of remnants after
expansion of Universe, see also \cite{ne}. They can be considered as some
windows through which the energy flows in our universe \cite{narl}, \cite%
{patwh}, \cite{sppwhite}, \cite{whr}. Such objects are probably unstable as
was argued in \cite{eard} (see also Sec. 15. 2 in \cite{fn}). However,
interest to them is still kept as they can lead to high energy collisions
between matter emerging from them and matter in our universe Particle
collisions under discussion can be a potential source of one more kind of
instability of white holes. Independently of whether or not astrophysical
white holes exist, consideration of physical processes in their vicinity is
of big methodical interest. (For instance, see recent work \cite{vis} and
references 43-65 therein.) General relativity is consistent with such
objects and, moreover, the full geodesically complete \ structure of
space-time includes both black and white hole regions that reveals itself
also in potential physical processes . Therefore, it is reasonable to study
them as fully as possible. Apart from this, such process in the geodesically
complete background can model more realistic ones when collapsing matter
collides with expanding one.

The paper is organized as follows. In Sec. \ref{set} we give the expression
for the metric, equations of motion of free particles and the main formulas
for particle collisions. In Sec. \ref{1} and \ref{fut} we describe in detail
main scenarios of particle collisions discussed in the present paper. In
Sec. \ref{3} we mention an additional scenario that incorporates features of
Scenarios 1 and 2. In Sec. \ref{kin} we discuss the relations between the
principle of kinematic censorship and properties of trajectories in
question. In Sec. \ref{dis} we discuss the properties of the Schwarzschild
time for a particle moving near a white whole and relate them to our
scenarios. In Sec. \ref{lemcoord} we describe these scenarios using the
expanding Lema\^{\i}tre frame. In Sec. \ref{sing} we consider processes with
particles near the singularity. In Sec. \ref{sum} we give summary of the
results obtained.

\section{General setup\label{set}}

Let us consider the static spherically symmetric metric

\begin{equation}
ds^{2}=-dt^{2}f+f^{-1}dr^{2}+r^{2}d\omega ^{2}\text{.}  \label{met}
\end{equation}%
If $f=1-\frac{r_{+}}{r}$, this corresponds to the Schwarzschild metric with $%
r_{+}$ being the horizon radius.

We restrict ourselves by pure radial motion. Choosing the plane in which a
particle moves to be $\theta =\frac{\pi }{2}$, we have equations of motion%
\begin{equation}
\dot{t}=\frac{\varepsilon }{f},  \label{t}
\end{equation}%
\begin{equation}
\dot{r}=\sigma P\text{, }P=\sqrt{\varepsilon ^{2}-f}.  \label{r}
\end{equation}%
Here, \ dot denotes derivative with respect to the proper time $\tau
,\varepsilon =\frac{E}{m}$, where $E$ is the Killing energy, $m$ being the
particle's mass, $\sigma =+1$ for motion in the outward direction and $%
\sigma =-1$ for for motion in the inward one.

At the horizon, the metric coefficients in the coordinate frame (\ref{met})
spoil. To repair this, there exists a number of other frames. We will stick
to the Kruskal-Szekeres coordinates (see, e.g. the textbook \cite{mtw},
Chapter VIII. 31. 4 - 31.5. A reader should bear in mind that we use another
notations). The concrete form of transformation from original coordinates to
Kruskal-Szekeres ones depends on the region under consideration.

In what follows, we will be interested in the energy $E_{c.m.}$ in the
center of mass frame of two colliding particles. By definition,%
\begin{equation}
E_{c.m.}^{2}=-(m_{1}u_{1\mu }+m_{2}u_{2\mu })(m_{1}u_{1}^{\mu
}+m_{2}u_{2}^{\mu })=m_{1}^{2}+m_{2}^{2}+2m_{1}m_{2}\gamma \text{,}
\label{ecm}
\end{equation}%
where $\gamma =-u_{1\mu }u_{2}^{\mu }$ is the Lorentz factor of relative
motion. If particles 1 \ and 2 move in the same plane, it follows from (\ref%
{t}), (\ref{r}) that%
\begin{equation}
\gamma =\frac{\varepsilon _{1}\varepsilon _{2}-\sigma _{1}\sigma
_{2}P_{1}P_{2}}{f}.  \label{ga}
\end{equation}

Hereafter, we will use both the original Schwarzschild coordinates (\ref{met}%
) and Kruskal-Szekeres ones. The whole space-time contains four regions
(see. e.g. Fig. 31.3 of \cite{mtw}). However, for our purposes we need to
consider only two of them. These are our region outside the horizon and the
white hole one $T^{+}$ under the horizon \cite{fn}.

\section{Scenario 1. Particle 1 in our region, particle 2 from white hole
region. Collision near the past \ horizon \label{1}}

We consider a scenario in which particle 2 emerges from the white hole
region and meets particle 1 in our region outside the horizon. See Fig. 1.
We assume that collision occurs in the point $r=r_{c}$ somewhere near the
right branch of the past horizon, so in this point $V_{c}\approx 0$ while $%
U_{c}<0$ is arbitrary.

\begin{figure}
    \centering
    \includegraphics[width=1\linewidth]{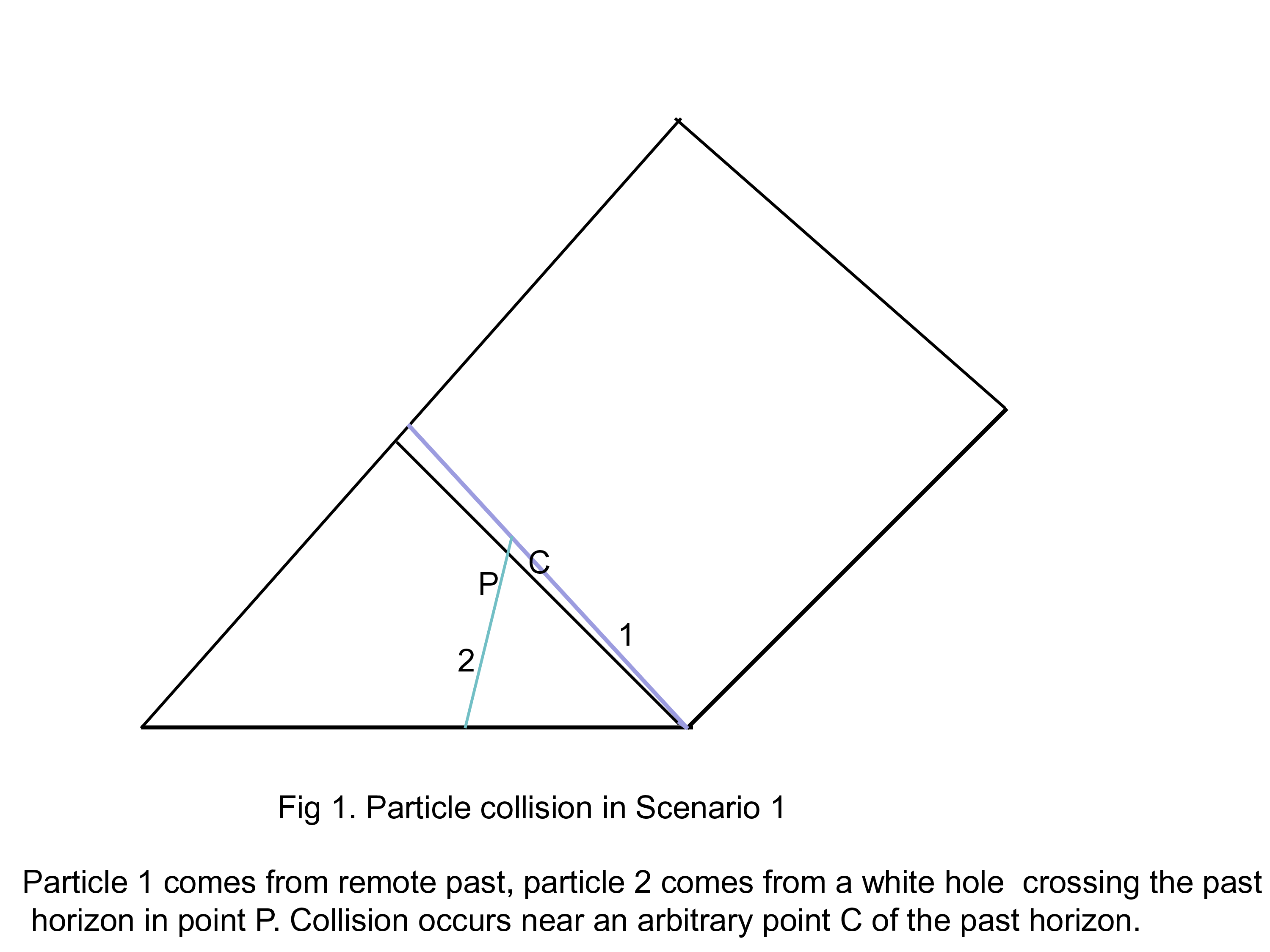}
\end{figure}

\subsection{Properties of the trajectory of particle 1}

Let us consider movement of a particle in region $r>r_{+}$. Then, the
relevant transformation from coordinates (\ref{met}) to Kruskal-Szekeres
ones is

\begin{equation}
U=-\exp (-\kappa u)\text{, }u=t-r^{\ast },  \label{U1}
\end{equation}%
\begin{equation}
V=\exp (\kappa v)\text{, }v=t+r^{\ast }\text{,}  \label{V1}
\end{equation}%
\begin{equation}
r^{\ast }=\int_{r_{1}}^{r}\frac{dr^{\prime }}{f(r^{\prime })},
\end{equation}%
where $r_{1}$ is some constant separated from $r_{+}$. Here, $\kappa $ is
the surface gravity$.$ For the Schwarzschild black hole, $\kappa =\frac{1}{%
2r_{+}}$ .

We can adjust a constant $r_{1}$ in such a way that%
\begin{equation}
r^{\ast }=r+r_{+}\ln \left\vert \frac{r-r_{+}}{r_{+}}\right\vert .
\label{tort}
\end{equation}%
Eqs. (\ref{U1}), (\ref{V1}) can be rewritten in the form%
\begin{equation}
U=-\sqrt{\frac{r}{r_{+}}-1}\exp (\frac{r-t}{2r_{+}})\text{,}  \label{U1r}
\end{equation}%
\begin{equation}
V=\sqrt{\frac{r}{r_{+}}-1}\exp (\frac{r+t}{2r_{+}}).  \label{V1r}
\end{equation}%
Let a particle move towards the horizon, so $\sigma =-1$ in (\ref{r}). Then,
it follows from (\ref{t}), (\ref{r}) that along the trajectory,%
\begin{equation}
\frac{dt}{dr}=-\frac{\varepsilon }{Pf},  \label{traj}
\end{equation}%
whence%
\begin{equation}
t=t_{0}^{(1)}+\varepsilon \int_{r}^{r_{0}}\frac{dr^{\prime }}{P(r^{\prime
})f(r^{\prime })}\text{,}  \label{ttr}
\end{equation}%
\begin{equation}
u=u_{0}^{(1)}+I_{+}(r,r_{0})\text{, }I_{+}(r,r_{0})=\int_{r}^{r_{0}}dr^{%
\prime }\frac{\varepsilon +P}{Pf},  \label{u}
\end{equation}%
\begin{equation}
v=v_{0}^{(1)}+I_{-}(r,r_{0})\text{, }I_{-}(r,r_{0})=\int_{r}^{r_{0}}\frac{%
\varepsilon -P}{Pf}dr^{\prime }\text{,}  \label{v}
\end{equation}%
where $t_{0}$, $r_{0}$, $u_{0}$ and $v_{0}$ are constants. Here, $r_{0}$
corresponds to the point where particle 1 starts its motion at the initial
moment $t_{0}$, $r\leq r_{0}$, the constants are related to each other
according to%
\begin{equation}
u_{0}^{(1)}=t_{0}^{(1)}+\int_{r_{0}}^{r_{1}}\frac{dr^{\prime }}{f(r^{\prime
})}\text{,}  \label{u0}
\end{equation}%
\begin{equation}
v_{0}=t_{0}-\int_{r_{0}}^{r_{1}}\frac{dr^{\prime }}{f(r^{\prime })}\text{,}
\label{v0}
\end{equation}%
where $u_{0}=t_{0}=v_{0}$, if $r_{0}=r_{1}$.

It is also instructive to write down equations of motion in the form%
\begin{equation}
\frac{dU}{dr}=\frac{\kappa U}{fP}(\varepsilon +P)\text{,}  \label{dU}
\end{equation}%
\begin{equation}
\frac{dV}{dr}=-\kappa V\frac{(\varepsilon -P)}{Pf}\text{.}  \label{dV}
\end{equation}

As $U_{c}<0$ is arbitrary according to assumption, $u_{c}$ is finite
(hereafter, subscript "c" means that the corresponding quantity refers to
the point of collision). Then, according to (\ref{V1}), $v_{c}$ should be
large negative when $r\rightarrow r_{+}$. We assume that $r_{0}$ is fixed
and finite, so $r_{0}^{\ast }$ is finite as well. Meanwhile, $t_{0}^{(1)}$
is a parameter allowed to vary.

If $t_{0}$ is finite, $u_{0}$ is finite as well. As the integral in (\ref{u}%
) diverges logarithmically when $r\rightarrow r_{+}$, it means that $%
u\rightarrow +\infty ,U\rightarrow 0$ and the point of collision is close to
the future horizon instead of being close to the past one, contrary to
assumption. To repair this, we assume that $u_{0}^{(1)}$ is big and negative
to compensate the integral which is big positive. This is achieved by the
choice of big negative $t_{0}^{(1)}$, so formally%
\begin{equation}
t_{0}^{(1)}\rightarrow -\infty \text{.}  \label{t0}
\end{equation}%
For $r$ close to the horizon,%
\begin{equation}
f\approx 2\kappa (r-r_{+})\text{.}  \label{fk}
\end{equation}

Then, we choose%
\begin{equation}
u_{0}^{(1)}=\frac{1}{\kappa }\ln (r_{c}-r_{+})+u_{1}  \label{u01}
\end{equation}%
where $u_{1}$ is some other finite constant. As a result, in the limit $%
r_{c}\rightarrow r_{+}$ we see that $u$ remains finite, so $U\rightarrow
U_{c}<0$ separated from zero. As far as $v_{c}$ is concerned, the integral
in (\ref{v}) remains finite for $r\rightarrow r_{+}\,.$ In doing so, $%
v_{c}\approx v_{0}+I_{-}(r_{+},r_{0})$. Taking into account (\ref{t0}), we
see that $v_{0}\rightarrow -\infty $, $v_{c}\rightarrow -\infty $, $%
V_{c}\rightarrow 0$, while $t_{c}$ is finite. Thus in the scenario under
discussion $U_{c}=O(1)<0$, and we are indeed near the right branch of the
past horizon. In doing so,%
\begin{equation}
V_{c}\approx \exp (\frac{t_{0}^{(1)}}{r_{+}})V^{\ast }\text{,}  \label{vc}
\end{equation}%
where the quantity $V^{\ast }$ remains finite nonzero in the limit when $%
r_{c}\rightarrow r_{+}$. Thus $V_{c}\approx 0$ due to (\ref{t0}).

If collision did not occur, particle 1 would have continued its motion and
hit the future horizon $U=0$ in the point with very small $V\sim V_{c}$.
Thus it would pass very close to the bifurcation point $U=V=0$.

To conclude this subsection, let us make some remarks of general character.
Usually, properties of particle motion are insensitive to the choice of the
moment $t_{0}$ of time start. The more so, this seems to be irrelevant in
the static metric However, now we deal with the process in which two
particles are involved, so that their trajectories should agree with each
other. This makes in eq. (\ref{ttr}) not only the $r$ dependence of the
integral important but also the choice of $t_{0}$. We will see below that
behavior of the constant of integration is an essential ingredient of high
energy collision scenarios. And, this is valid not only for scenario 1 but
for scenario 2 as well.

\subsection{Properties of the trajectory of particle 2}

Particle 2 moves emerges from a white hole. In this region, the
Kruskal-Szekeres transformation reads

\begin{equation}
U=-\exp (-\kappa u)\text{, }u=t-r^{\ast },
\end{equation}%
\begin{equation}
V=-\exp (\kappa v)\text{, }v=t+r^{\ast }\text{.}
\end{equation}

Particle 2 should hit the right branch of the past horizon in the point with 
$U<0$, $V=0$. Therefore, we must have $u_{0}$, $v_{0}$ for this particle
finite.

As it moves from smaller $r$ to bigger $r$, we have $\dot{r}_{2}>0$.
Meanwhile, as particle 1 moves towards the horizon, $\dot{r}_{1}<0$. Thus $%
\sigma _{1}\sigma _{2}=-1$ and near the horizon we have from (\ref{ga})$%
\backsim $%
\begin{equation}
\gamma \approx \frac{2\varepsilon _{1}\varepsilon _{2}}{f}  \label{ga2}
\end{equation}%
that diverges in the limit $r\rightarrow r_{+}$, $f\rightarrow 0$.

It is instructive to rewrite this formula in the form that explicitly
includes the parameter $t_{0}$. It follows from (\ref{U1}), (\ref{V1}) that%
\begin{equation}
UV=-\exp (\frac{r}{r_{+}})\frac{(r-r_{+})}{r_{+}},  \label{uv}
\end{equation}%
where we used (\ref{tort}) and the value $\kappa =\frac{1}{2r_{+}}$ for the
Schwarzschild metric. Near the horizon, (\ref{fk}) gives us%
\begin{equation}
f\approx -UV\text{.}  \label{fuv}
\end{equation}%
Eqs. (\ref{uv}) and (\ref{fuv}) are general expressions valid independently
of any scenario of collision.

If we consider collisions and assume Scenario 1, we must take into account (%
\ref{vc}) and the fact that $U_{c}=O(1)$. Then,

\begin{equation}
r_{c}-r_{+}\backsim \exp (\frac{t_{0}^{(2)}}{r_{+}})\backsim V_{c}\text{.}
\end{equation}

\begin{equation}
\gamma \backsim V_{c}^{-1}\backsim \exp (-\frac{t_{0}^{(2)}}{r_{+}})\backsim 
\frac{1}{r_{c}-r_{+}}\text{,}
\end{equation}

Thus we see that the same parameter $t_{0}^{(2)}$ (which is large negative)
governs both the behavior of the Lorentz gamma factor (hence, \thinspace
\thinspace $E_{c.m.}$) and the proximity of the point of collision to the
horizon. To distinguish between particles, we indicate explictly their
number in a superscript.

\section{Scenario 2. Particle 1 in our region, particle 2 from white hole
region. Collision near the future horizon \label{fut}}

Now, we consider somewhat different scenario. Particle 2 comes from the
white hole region and collides with particle 1 near the future horizon in
region I. See Fig. 2. In our region we can use expressions (\ref{U1}), (\ref%
{V1}). In contrast to the scenario described in Sec. \ref{1}, where in the
point of collision $U_{c}<0$ is arbitrary and $V_{c}\rightarrow 0$, now $%
U_{c}\rightarrow 0$ and $V_{c}>0$ is arbitrary, $t_{c}\rightarrow +\infty $.

\begin{figure}
    \centering
    \includegraphics[width=1\linewidth]{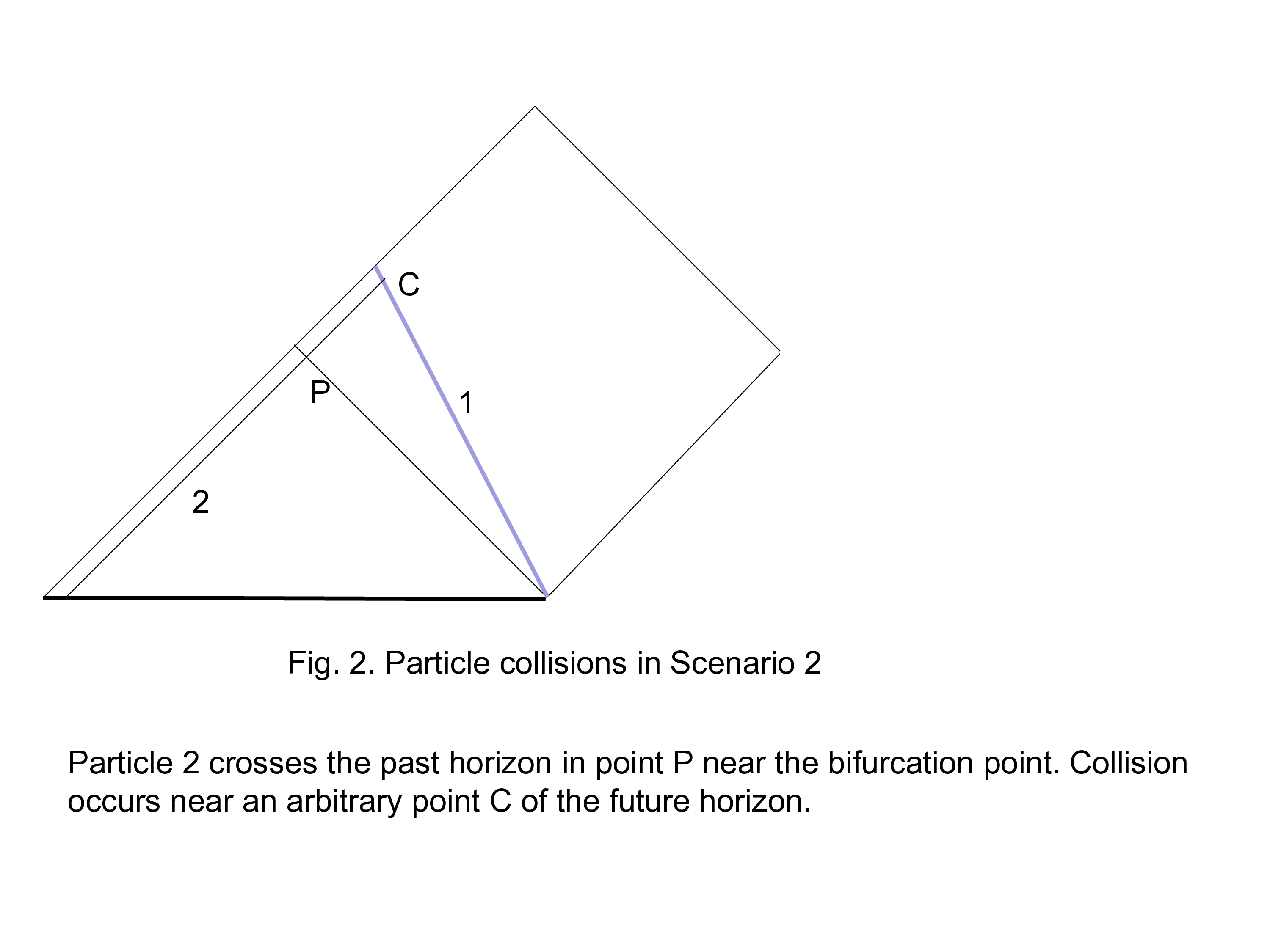}
\end{figure}

\subsection{Properties of the trajectory of particle 1}

For particle 1, this can be achieved if both $u_{0}$ and $v_{0}$ in (\ref{u}%
), (\ref{v}) are finite. In this sense, there are no special restrictions on
this trajectory. In doing so, $t_{c}\rightarrow +\infty $.

\subsection{Properties of the trajectory of particle 2}

Let us consider now particle 2. In the scenario in question it crosses the
past horizon and further moves in the outward direction, so $\sigma
_{2}=+1\, $, and instead of (\ref{traj}), we have along the trajectory of
particle 2%
\begin{equation}
\frac{dt}{dr}=+\frac{\varepsilon }{Pf}  \label{tr2}
\end{equation}%
and%
\begin{equation}
t=t_{0}^{(2)}-\varepsilon \int_{r}^{r_{0}}\frac{dr^{\prime }}{P(r^{\prime
})f(r^{\prime })}\text{,}  \label{t2}
\end{equation}%
\begin{equation}
u=u_{0}^{(2)}+\int_{r}^{r_{0}}dr^{\prime }\frac{P-\varepsilon }{Pf},
\label{u2}
\end{equation}%
\begin{equation}
v=v_{0}^{(2)}-\int_{r}^{r_{0}}\frac{\varepsilon +P}{Pf}dr^{\prime }\text{,}
\label{v2}
\end{equation}%
where 
\begin{equation}
u_{0}^{(2)}=t_{0}^{(2)}+\int_{r_{0}}^{r_{1}}\frac{dr^{\prime }}{f(r^{\prime
})}\text{,}
\end{equation}%
\begin{equation}
v_{0}^{(2)}=t_{0}^{(2)}-\int_{r_{0}}^{r_{1}}\frac{dr^{\prime }}{f(r^{\prime
})}\text{.}
\end{equation}%
In terms of the Kruskal-Szekeres coordinates, for particle 2 we have in the
region outside the horzon%
\begin{equation}
\frac{dU}{dr}=-\frac{\kappa U}{fP}(\varepsilon -P)\text{,}  \label{dU2}
\end{equation}%
\begin{equation}
\frac{dV}{dr}=\frac{\kappa V}{fP}\left( \varepsilon +P\right) .  \label{dV2}
\end{equation}%
We want to adjust the trajectory of particle 2 in such a way that near the
future $\ $horizon $V_{c}>0$ is arbitrary, $U_{c}\approx 0$. It is seen from
(\ref{u2}) that the integral remains finite in the limit $r\rightarrow r_{+}$%
, $f\rightarrow 0$ whereas the integral in (\ref{v2}) tends to $+\infty $.
Therefore, the only way to achieve the aforementioned properties is to take $%
u_{0}^{(2)}$ and $v_{0}^{(2)}$ big positive. Then, when particle 2 crosses
the past horizon, $V=0$ and $U=U_{p}<0$ with $\left\vert U_{p}\right\vert
\ll 1$. Meanwhile, when it approaches $r_{c}$, both terms in (\ref{v2})
compensate each other and $V_{c}=O(1)$. The fact that $\left\vert
U_{p}\right\vert \ll 1$ means that particle 2 crosses the past horizon \
near the bifurcation point $U=0=V$.

More precisely, it is sufficient to choose 
\begin{equation}
v_{0}^{(2)}=\frac{1}{\kappa }\left\vert \ln \frac{r_{c}-r_{+}}{r_{+}}%
\right\vert +v_{2}\text{,}  \label{v02}
\end{equation}%
where $v_{2}$ is some unimportant constant. In doing so, particle 2 crosses
the past horizon $r=r_{+}$ in the point with%
\begin{equation}
\left\vert U_{p}\right\vert \backsim \exp (-\frac{t_{0}^{(2)}}{r_{+}}%
)\backsim r_{c}-r_{+}.
\end{equation}

In the limit under discussion both $t_{0}^{(2)}$, $v_{0}^{(2)}\rightarrow
+\infty $, and $t_{c}^{(2)}\rightarrow +\infty $ as well. In doing so, $%
v_{c} $ \ remains finite. Thus collision occurs in infinite future (from the
viewpoint of a remote observer in our world) as it should be.

As now in the point of collision $\dot{r}_{1}<0$ and $\dot{r}_{2}>0,$ the
Lorentz factor (\ref{ga2}) is unbounded. It can be rewritten in the form%
\begin{equation}
\gamma \backsim \frac{1}{r_{c}-r_{+}}\backsim \frac{1}{\left\vert
U_{p}\right\vert }\text{.}
\end{equation}

The same parameter $t_{0}^{(2)}$ governs the behavior of $\gamma $ (hence $%
E_{c.m.}$) in the point of collision $r_{c}$ and the proximity of the
trajectory of particle 2 to the bifurcation point when it crossed the past
horizon before collision.

Geometrically, it is convenient to interpret the situation in terms of
coordinates $r=-T$, $t=y$ under the horizon \cite{nov61},%
\begin{equation}
ds^{2}=-\frac{dT^{2}}{g}+gdy^{2}+T^{2}d\omega ^{2}\text{,}  \label{nov}
\end{equation}%
where $g=-f\geq 0$. \ The surface of a constant $r$ looks like a
hypercylinder. Then, the condition $t_{0}^{(2)}\rightarrow +\infty $ is
translated to $y_{0}^{(2)}\rightarrow +\infty $ for particle 2 where $%
y_{0}^{(2)}$ is an initial value of $y$. Thus particle 2 should start its
motion very far on the leg of the aforementioned hypercylinder.

\begin{figure}
    \centering
    \includegraphics[width=1\linewidth]{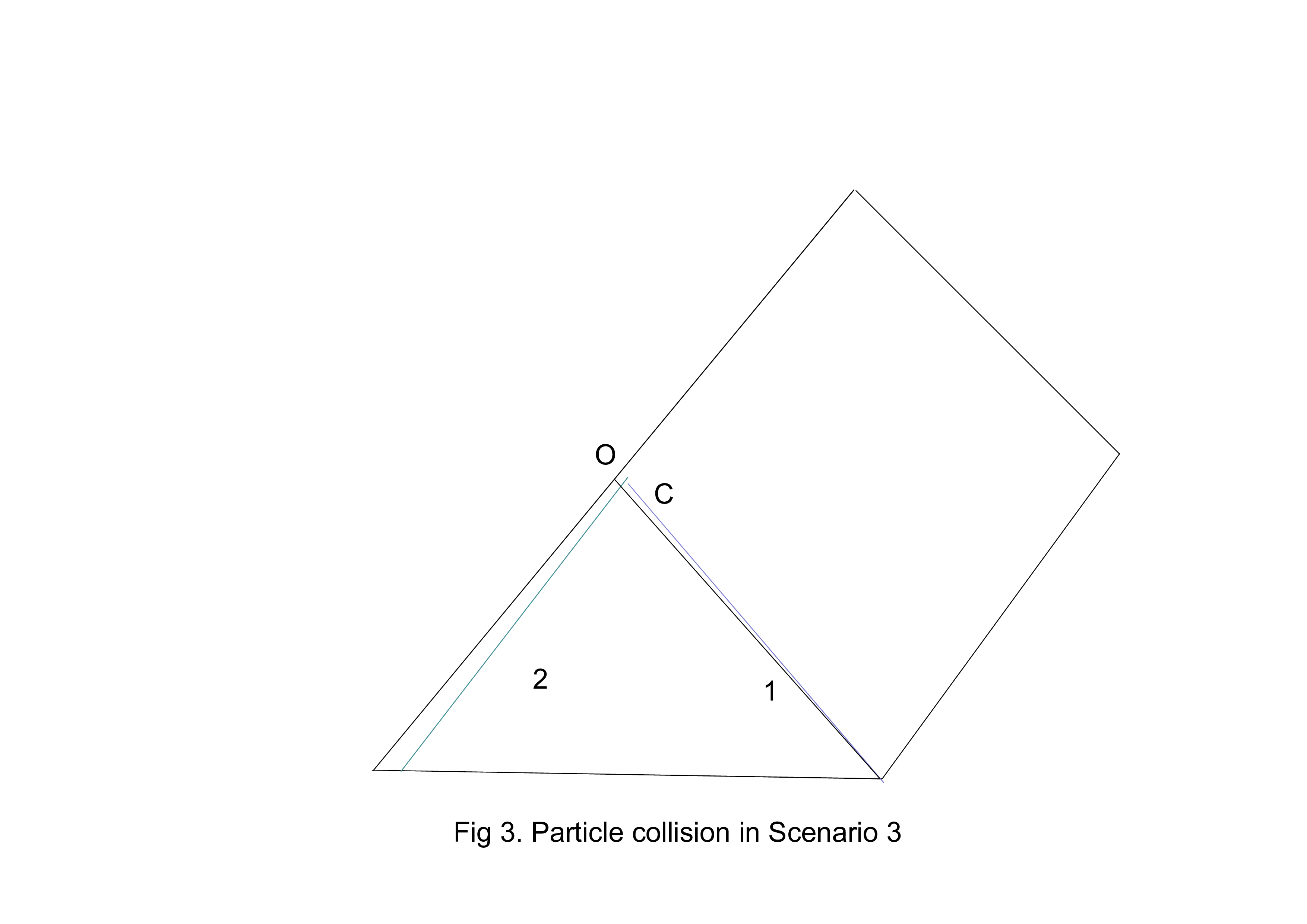}
\end{figure}

\section{Scenario 3 \label{3}}

Scenarios 1 and 2 look complementary to each other since in the first case $%
t_{0}^{(1)}\rightarrow -\infty $, $t_{0}^{(2)}$ is finite and in the second
case $t_{0}^{(1)}$ is finite and $t_{0}^{(2)}\rightarrow +\infty $.
Meanwhile, there is a way to combine them in another scenario. This can be
achieved if simultaneously $t_{0}^{(1)}\rightarrow -\infty $ and $%
t_{0}^{(2)}\rightarrow +\infty $. See Fig. 3. In the Carter-Penrose diagrams
trajectories of both particles go closely to two branches of the past
horizon.

\section{Divergent $E_{c.m.}$ , unattainability of the speed of light and
kinematic censorship \label{kin}}

We would like to stress some meaningful features of the scenarios under
discussion that are connected with manifestation of kinematic censorship.

\subsection{Scenario 1}

For any finite $t_{0}^{(1)}$, the energy $E_{c.m.}$ remains finite, although
it can be made as big as one likes. In the limit $t_{0}^{(1)}\rightarrow
-\infty $ it diverges. But, clearly, the moment $t_{0}$ cannot be equal to
infinity literally. To some extent, this example is complimentary to that
for extremal black holes. For them, $E_{c.m.}$ is formally infinite if
collision would happen precisely on the horizon and one of colliding
particles is fine-tuned. But this requires infinite proper time $\tau $ to
reach the horizon \cite{ted}. We can take it not exactly fine-tuned, then $%
\tau $ is finite. But $E_{c.m.}$ is finite as well. In our example it is
coordinate (not proper) time which diverges and this restricts $E_{c.m.}$

There is also another interesting property of this scenario. As is mentioned
above, without collision particle 1 would continue its trajectory and hit
the future horizon in the point very close to the bifurcation point. The
proximity to it is governed by the same parameter $t_{0}^{(1)}$. Formally,
infinite $\left\vert t_{0}^{(1)}\right\vert $ would mean that particle 1
passes though the bifurcation point precisely. In turn, this would mean
movement along the horizon. But a massive particle cannot move along the
light-like surface, its trajectory is time-like. Thus we can say that
impossibility to reach infinite $E_{c.m.}$ is connected (although
indirectly) with impossibility for a massive particle to attain the speed of
light.

We can rephrase this in somewhat different way. We arrange collision in a
point with a given $t_{c}$.If we keep $t_{c}$ fixed, when sending $%
t_{0}^{(1)}\rightarrow -\infty $, the point of collision becomes closer and
closer to the horizon, the integral in eq. (\ref{ttr}) grows without bound
in such a way that divergent terms compensate each other. Meanwhile, in the
movement in the static background (see, e.g. eq. 88.9 in \cite{LL}) the
conserved energy%
\begin{equation}
E=\frac{m\sqrt{f}}{\sqrt{1-V^{2}}}\text{,}
\end{equation}%
where $V$ is velocity measured by a static observer, the speed of light $c=1$%
. As particle 1 falls towards the horizon, the value of $r$ decreases from $%
r>r_{+}$ to $r=r_{+}$ and the horizon is reachable only in the limit. In
doing so, $\dot{r}<0$ when crossing the horizon, so it cannot move along $%
r=const=r_{+}$. In other words, before reaching the future horizon, a
particle cannot move along the past horizon. Correspondingly, collision with
particle 2 cannot occur exactly on the past horizon and $\ E_{c.m.}$ cannot
become infinite in agreement with the principle of the kinematic censorship.
The limit $V=1$ would have been achieved for a massive particle on the
future (not past) horizon, if collision had not occured. Thus three
circumstances cannot be realized for a massive particle: (i) infinite $%
E_{c.m.}$, (ii) collision on the past horizon, (iii) exact equality $V=1$.

\subsection{Scenario 2}

In this scenario, for particle 2 the quantity $t_{0}^{(2)}\rightarrow
+\infty $ along with infinite $E_{c.m.}.$ Under the horizon, as is said
above, this implies infinite distance along the leg of cylinder that is
impossible, distance can be as large as one likes but it remains finite
anyway. Also, the same arguments as above, show that infinite $t_{0}^{(2)}$
would mean that a massive particle moves along the light-like surface. The
difference between scenarios 1 and 2 consists in this context in that in the
first case the past horizon is relevant whereas in the second case it is the
future horizon.

The above discussion shows relationship between changing the constants $%
t_{0}^{(1)}$, $t_{0}^{(2)}$and the proximity of collision to the horizon,
hence an infinite growth of $E_{c.m.}$ due to velocity $V\,$\ approaching 1.
By itself, the fact that from the viewpoint of a stationary observer $%
V\rightarrow 1$ in the horizon limit, is well known. But now, we put it in
the context directly connected with the principle of kinematic censorship.
One should bear \ in mind that the static frame in which $V\rightarrow 1$
becomes singular on the horizon where it cannot be realized by physical
bodies. The question how these issues look for a physical observers
measuring velocities is \ considered below in Sec. \ref{lemcoord}. In this
sense, we trace the examination of the kinematic censorship from two
different points.

\section{Discussion: collision near white hole and behavior of the
Schwarzschild time\label{dis}}

In the above consideration, we were faced with quite unusual behavior of the
Schwarzschild time in Scenario 1: although collision happens near the
horizon, $t_{c}$ remains finite and does not diverge when $r_{c}\rightarrow
r_{+}$. This is contrasted with Scenario 2, where a quite standard property
well known from textbooks reveals itself: when a particle approaches a black
hole, its time $t$ measured by a remote observer, diverges. To elucidate
corresponding subtleties, let us get back to the calculation of $t_{c}$. It
is instructive to consider this for each of two particles separately.

\subsection{Particle 1}

For particle 1, eq. (\ref{ttr}) with (\ref{u0}), (\ref{v0}), (\ref{u01})
taken into account, gives us that the big negative divergent $t_{0}$
compensates the big positive integral in such a way that the net outcome is
finite in the limit $r_{0}\rightarrow r_{+}$. Thus $t_{c}$ remains finite.
We would like to stress that this conclusion has nothing to do with the
existence of a white hole region under the past horizon and is valid
independently of whether or not we consider collisions. However, it relies
heavily on distinction between the future horizon and the past horizon. In
the first case, when a particle approaches the future horizon, i.e. falls in
a \ black hole, $t\rightarrow \infty .$

\subsection{Particle 2}

It is instructive to look for the trajectory of particle 2 even in a more
wide context. Let us trace how $t^{(2)}(r)$ changes along the trajectory of
particle 2. In the region outside the horizon $t^{(2)}(r)$ is finite for any 
$r>r_{+}$. This allows us to arrange collision between particles 1 and 2 in
some point $r_{c}>r_{+}$, so that $t^{(2)}(r_{c})=t^{(1)}(r_{c})$.
Meanwhile, it follows from (\ref{tr2}) that along the trajectory 
\begin{equation}
t^{(2)}(r)=t_{0}^{(2)}-\int_{r}^{r_{c}}\frac{\varepsilon _{2}dr}{P(r^{\prime
})f(r^{\prime })}\text{,}  \label{ts2}
\end{equation}%
where $t_{0}^{(2)}=t^{(2)}(r_{c})$. When we trace behavior of $t^{(2)}(r)$
back in time and $r\rightarrow r_{+}$, $t^{(2)}(r)\rightarrow -\infty $. In
other words, a particle crossed the past horizon in an infinite past.

The situation is opposite to that for ingoing particle 1 for which $t(r)$ is
finite for any $r>r_{+}$ but $t(r)\rightarrow +\infty $ when $r\rightarrow
r_{+}$.

The contents of the present subsection is a counterpart of pages 114 -\ 115
of \cite{zn1}, where appearance of massless particles from a white hole was
discussed. It was explained there that an observer in the outer world is
able to see all interior of a white hole starting from the singularity. In
somewhat different language, we argued above that a similar phenomenon is
valid for massive particles. An observer residing at any $r>r_{+}$ is able
to catch particles coming from the interior and, as a consequence, to see
all the interior of a white hole including the singularity.

Thus it is quite typical for outgoing particle 2 to have $t_{c}$ finite.
What is more unusual is that $t_{c}$ remains finite in the limit $%
r_{c}\rightarrow r_{+}$ for ingoing particle 1. It would seem that this is
contrasted with the behavior known even from textbooks according to which $%
t_{c}\rightarrow \infty $ in such circumstances. However, one should bear in
mind two qualitative features here: (i) now the horizon is a past horizon,
not a future one, (ii) collision occurs near a white (not a black) hole.
Mathematically, finite $t_{c}$ appears due to compensation of a big constant 
$t_{0}^{(1)}$ and big (almost diverging) integral in (\ref{ttr}) as is
explained above in \ref{1}. And, in any case the difference in time $%
t_{c}-t_{0}$ diverges when a point of collision approaches the horizon.

\section{How does the picture look like in the Lema\^{\i}tre coordinates? 
\label{lemcoord}}

Apart from the Kruskal-Szekeres coordinates, there exists one more popular
frame that remains regular near the event horizon. This is the Lema\^{\i}tre
frame. It is quite instructive to describe in this frame particle collisions
under discussion. The connection between the Lema\^{\i}tre frame and the
original curvature coordinate is well known and is described in textbooks
(see, e.g. Sec. 102 in \cite{LL}). Here, one can distinguish between two
different cases - the contracting Lema\^{\i}tre frame and the expanding one.
The contracting Lema\^{\i}tre frame is composed by particles that fall
freely from infinity with $\varepsilon =1$. The expanding Lema\^{\i}tre
frame is composed by particles with $\varepsilon =1$ that move from the
singularity, cross the white hole region and enter our part of space. In our
context, where the presence of a white hole is an essential ingredient, it
is the expanding one that is preferable since it encompasses both the white
hole region ($T^{+}$ one according to the terminology of \cite{fn}) and the
outer one ($R$ region). Meanwhile, the contracting Lema\^{\i}tre frame is
unable to describe the history of a particle in the $T^{+}$ region.

In the expanding Lema\^{\i}tre frame we introduce new coordinates%
\begin{equation}
\hat{t}=t-\int_{r_{c}}^{r}\frac{dr\sqrt{1-f}}{f}\text{,}  \label{lems}
\end{equation}%
where we choose the constant of integration such that $\hat{t}=t$ at $%
r=r_{c},$%
\begin{equation}
\rho =-t+\int_{r_{c}}^{r}\frac{dr}{f\sqrt{1-f}},
\end{equation}%
in which the metric reads%
\begin{equation}
ds^{2}=-d\hat{t}^{2}+(1-f)d\rho ^{2}+r^{2}d\omega ^{2}\text{.}  \label{lem}
\end{equation}

Along the trajectory of an outgoing particle 2, we have relations (\ref{tr2}%
), (\ref{ts2}), and%
\begin{equation}
\hat{t}=t_{0}^{(2)}+\varepsilon \int_{r_{c}}^{r}\frac{dr^{\prime }}{%
f(r^{\prime })\sqrt{\varepsilon ^{2}-f}}-\int_{r_{c}}^{r}\frac{dr\sqrt{1-f}}{%
f}\text{.}  \label{p2e}
\end{equation}

Near the horizon when $r_{c}\rightarrow r_{+}+0$, the metric function $%
f\rightarrow 0$ and both integrals in (\ref{p2e}) compensate each other, so $%
\hat{t}_{c}$ is finite for any finite $t_{0}^{(2)}$. This is quite natural
as this frame is designed just for description of outgoing particles. In
other words, divergent $t_{c}(r)$ for $r\rightarrow r_{+}$ is combined with
the finiteness of $\hat{t}(r)$, as it should be. This agrees with the
behavior of $t(r)$ described in Sec. \ref{dis}.

For an ingoing particle 1, 
\begin{equation}
t=t_{0}^{(1)}-\varepsilon \int_{r_{0}}^{r}\frac{dr^{\prime }}{f(r^{\prime
})P(r^{\prime })}
\end{equation}%
and 
\begin{equation}
\hat{t}=t_{0}^{(1)}-\varepsilon \int_{r_{0}}^{r}\frac{dr^{\prime }}{%
f(r^{\prime })\sqrt{\varepsilon ^{2}-f}}-\int_{r_{0}}^{r}\frac{dr\sqrt{1-f}}{%
f}\text{.}  \label{in}
\end{equation}

As particle 1 starts at $r=r_{0}>r_{+}$, both integrals have the same sign
and diverge when $r\rightarrow r_{+}.$ More aspects of relation between
different frames and particles moving in the background under discussion can
be found in \cite{we2}. The definition of constants in (\ref{lems}) and (\ref%
{in}) for particles 1 and 2 do not coincide. This is quite natural since
their trajectories are qualitatively different. Tracing back in time, an
outgoing particle 2 approaches $r_{c}$ slightly above the horizon, an
ingoing particle 1 approaches $r_{0}$.

Let us consider the same scenarios as before.

\subsection{Scenario 1}

Particle 2 appears from the white hole region and moves in the outward
direction where it meets particle 1 close to the past horizon. Particle 1 is
ingoing. For it, (\ref{in}) holds. Particle 2 is outgoing, so (\ref{p2e})
holds. We want both particles to meet, so $\hat{t}_{1}=\hat{t}_{2}$. To
achieve this, we must adjust constants $t_{0}^{(1)}$, $t_{0}^{(2)}$. Then,
for collisions near the horizon, \thinspace we take finite $t_{0}^{(2)}$ and 
$t_{0}^{(1)}\rightarrow -\infty $. Thus we again come to the same conclusion
obtained above in the analysis that uses Kruskal-Szekeres coordinates.

We see that there is full \ complementarity between pictures of particle
motion and collision in both frames. The ingoing particle is "unnatural"
with respect to the expanding Lema\^{\i}tre and has formally divergent Lema%
\^{\i}tre time when it starts its motion. Note, that collision itself occurs
for a finite Lema\^{\i}tre time even if the collision energy is unbounded.

It is known that the velocity of an ingoing particle with respect to
expanding Lema\^{\i}tre frame (as well as the velocity of an outgoing
particle with respect to contracting Lema\^{\i}tre frame) reaches the
velocity of light at the horizon \cite{we2} (especially see Table 2 there).
This leads us again to the connection between divergent collision energy and
a massive particle with the velocity of light. Indeed, the velocity with
respect to Lema\^{\i}tre frame can be measured directly by free moving
particle since this frame is a free moving frame. So that, an ordinary
particle crossing a white hole horizon outward at some time $\hat{t}_{c}$
would be able to measure a velocity of an ingoing particle colliding with it
at a horizon and get the embarassing result $v=c$! What prevents such a
situation is the fact that the expanding Lema\^{\i}tre time for an ingoing
particle to reach the horizon is infinite, so that it should start motion in
infinitely remote past to compensate diverging integral in (\ref{in}).

\subsection{Scenario 2}

Now, there are no special restriction imposed on particle 1, $t_{0}^{(1)}$
is finite. As a result, for the point of collision $\hat{t}\rightarrow
+\infty $. For particle 2, we choose $t_{0}^{(2)}\rightarrow +\infty $ to
achieve $\hat{t}_{1}=\hat{t}_{2}$. Note that in this scenario the Lema\^{\i}%
tre time of the collision diverges for divergent energy.

\section{Particle decay near singularity\label{sing}}

Up to now, we mainly considered particle collisions near the horizon.
Meanwhile, it is also of interest to consider an immediate vicinity of the
singularity. In doing so, instead of collisions, we will discuss more simple
case, when particle 0 decays to fragments 1 and 2. We will see that even in
this simplified settings, results are quite interesting and have nontrivial
consequences. In the present article we are concerned with processes outside
the horizon. However, it makes sense to add the decay in question into a
general scheme, provided we consider such a decay in the white hole region.
Motivation comes from the fact that after decay new particles can appear in
the outer region.

As we are interested in the region beyond the horizon, $f=-g<0$, it is
convenient to use the metric in the form (\ref{nov}).

As before, we assume that all angular momenta are zero. Then, equations of
motion for a free particle read%
\begin{equation}
m\dot{y}=-\frac{E}{g}\text{,}
\end{equation}%
\begin{equation}
m\dot{T}=\sqrt{E^{2}+gm^{2}}\text{,}
\end{equation}%
where $E$ is a constant. Now, the quantity $E$ has the meaning of momentum
(not energy) along the $y-$direction and $\,m\dot{T}$ has the meaning of
energy (up to the factor $g$). As now the metric is homogeneous and
non-static, it is $E$ which is conserved whereas the energy is not. However,
we assume that in the point of decay conservation laws hold:%
\begin{equation}
E_{0}=E_{1}+E_{2}\text{,}
\end{equation}%
\begin{equation}
\sqrt{E_{0}^{2}+gm_{0}^{2}}=\sqrt{E_{1}^{2}+gm_{1}^{2}}+\sqrt{%
E_{2}^{2}+gm_{2}^{2}}.
\end{equation}

After some algebra, one obtains%
\begin{equation}
E_{1}=\frac{b_{1}}{2m_{0}^{2}}E_{0}\pm \frac{\sqrt{d}\sqrt{%
E_{0}^{2}+m_{0}^{2}g}}{2m_{0}^{2}}\text{,}
\end{equation}%
\begin{equation}
E_{2}=\frac{b_{2}}{2m_{0}^{2}}E_{0}\mp \frac{\sqrt{d}\sqrt{%
E_{0}^{2}+m_{0}^{2}g}}{2m_{0}^{2}},
\end{equation}%
where%
\begin{equation}
b_{1,2}=m_{0}^{2}+m_{1,2}^{2}-m_{2,1}^{2}\text{,}
\end{equation}%
\begin{equation}
d=b_{1}^{2}-4m_{0}^{2}m_{1}^{2}=b_{2}^{2}-4m_{0}^{2}m_{2}^{2}.
\end{equation}

These formulas are similar to eqs. 45, 46 and 51, 52 of \cite{genpen} where
decay outside the horizon was considered. However, now they hold for a
non-static metric.

Near singularity $g\rightarrow \infty $. Then,%
\begin{equation}
E_{1,2}\sim \pm \frac{\sqrt{d}}{2m_{0}}\sqrt{g}
\end{equation}%
independently of $E_{0}$.

The quantity $E$ under the horizon has the meaning of momentum, not energy.
It conserves due to homogeneity of the metric in this region. When a
particle crosses the horizon, the value of this integral of motion remains
the same but the meaning changes: momentum "converts" into energy.
Therefore, if $E_{1,2}$ is indefinitely big in the point of decay, it
remains indefinitely big in the vicinity of the horizon from any side of it.
Thus if decay occurs inside a white hole region near the singularity,
particles with unbounded $E$ can cross the horizon and enter the outer $R$
region with the same unbounded $E$: particle with $E>0$ enters the right $R$
region and that with $E<0$ enters the left $R$ region. As a result, flux of
such particles can cause strong backreaction and destroy the past horizon.

Although detailed consideration in the black hole region is beyond the scope
of the present work, it is to the point to make short comparison. The
aforementioned destruction does not happen for the black hole horizon
(future horizon) since in the $T^{-}$ region (we follow classification
developed in \cite{novrt}) decay near the singularity occurs when particles
already crossed the horizon in the past. However, such strong backreaction
can destroy, in principle, the black hole singularity. Thus the effect of
decay in this case is very similar to that due to particle collisions near
the singularity \cite{rad}. In this sense, there is some complementarity:
decay near the white hole singularity seems to destroy the white hole
horizon, decay near the black hole singularity destroys the singularity
itself.

It is worth stressing that the results in question are obtained pure
classically and do not require resort to quantum effects \cite{qwhite}, \cite%
{waldwhite}.

\section{Summary and conclusions \label{sum}}

High energy collisions in the Schwarzschild metric are possible due to
participation of a particle coming from a white hole region that are absent
in the standard BSW effect. The main properties of such collisions are
summarized in Table I. We indicated in scenario 1 proximity of particle 1 to
the bifurcation point. However, the reservation is in order here that this
is "virtual" proximity for the trajectory which would continue without
collision with particle 2. In lines 1 and 2 we indicate the time of
collision. We would like to stress that the symbols $-\infty $ or $+\infty $
are pure conditional and used for the sake of brevity. They mean that in the
horizon limit $r_{c}\rightarrow r_{+}$ the corresponding quantity tends to
plus or minus infinity, although for any $r_{c}>r_{+}$ \ it is finite.
Scenario 3 combines features of those 1 and 2.

\begin{tabular}{|l|l|l|l|}
\hline
& Scenario 1 & Scenario 2 & Scenario 3 \\ \hline
$t_{c}$ & finite & $+\infty $ & $+\infty $ \\ \hline
$\hat{t}_{c}$ & finite & $+\infty $ & $+\infty $ \\ \hline
$t_{0}^{(1)}$ & $-\infty $ & finite & $-\infty $ \\ \hline
$t_{0}^{(2)}$ & finite & $+\infty $ & $+\infty $ \\ \hline
proximity to bifurcation point & particle 1 (virtually) & particle 2 & both
\\ \hline
\end{tabular}

Table I. Properties of constants $t_{0}^{(1)}$, $t_{0}^{(2)}$ and behavior
of the Lema\^{\i}tre time in the point of collision with unbounded energy $%
E_{c.m.}$.

We would like to stress that for particle 2 $t_{0}^{(2)}$ cannot be
identified as an initial moment. Under the horizon, where this particle
merges from, the variable $t$ has a spatial character, $t=y$, so its
divergent value means that it starts its motion very far along the leg of a
corresponding hypercilinder in the metric (\ref{nov}).

We traced in detail how the same conditions govern both the indefinite
growth of $E_{c.m.}$ of colliding particles, proximity of corresponding
events to the horizon and unattainability of a speed of light for a massive
particle. In this sense, we reveal physical meaning of kinematic censorship
for scenarios under discussion.

We also considered process of particle decay near the singularity and found
that it leads to indefinitely large energies of its debris. However,
literally infinite energy requires a decay exactly at a singularity where
known physics fails. This result is also obtained purely classically.

We descibed the processes under discussion in the Kruskal-Szekeres frames as
well as in the expanding Lema\^{\i}tre one. Such a frame is rather rarely
used in literature but now its using was motivated by a context since it is
this version of the Lema\^{\i}tre one that is able to describe history of
particles in the regions of interest.

It is reasonable to believe that unboundedness of the collision and decay
energies from above (despite that literal infinity cannot be achieved) can
be considered as a new argument against the existence of astrophysical white
holes, using pure classical arguments and including processes both near the
past horizon and near the white hole singularity. This issue needs further
investigations.

\section{Acknowledgement}

We recall with gratitude A. A. Starobinsky with whom we had time to discuss
this work before his sudden death.

\end{document}